\documentclass[a4paper,twocolumn,english,superscriptaddress,showpacs,aps]{revtex4}

\usepackage{amsmath}
\usepackage{amssymb}
\usepackage{graphicx}
\usepackage{color}
\usepackage{hyperref}

\begin{document}
\title{Pattern dynamics near homoclinic bifurcation in Rayleigh-B\'{e}nard convection}
\author{Pinaki Pal}
\affiliation{Department of Mathematics, National Institute of Technology, Durgapur-713 209, India}
\author{Krishna Kumar}
\author{Priyanka Maity}
\affiliation{Department of Physics and Meteorology, Indian Institute of Technology, Kharagpur-721 302, India} 
\author{Syamal Kumar Dana}
\affiliation{CSIR-Indian Institute of Chemical Biology, Jadavpur, Kolkata-700 032, India}

\begin{abstract}
We report for the first time the pattern dynamics in the vicinity of an inverse homoclinic bifurcation in an extended dissipative system. We observe, in  direct numerical simulations of three dimensional Rayleigh-B\'{e}nard convection, a spontaneous breaking of a competition of two mutually perpendicular sets of oscillating cross rolls to one of two possible sets of oscillating cross rolls as the Rayleigh number is raised above a critical value. The time period of the cross-roll patterns diverges, and shows scaling behavior near the bifurcation point. This is an example of a transition from nonlocal to local pattern dynamics near an inverse homoclinic bifurcation. We also present a simple four-mode model that captures the pattern dynamics quite well.
\end{abstract}

\pacs{47.20.Ky, 47.55.pb, 47.20.Bp}

\maketitle

Extended dissipative systems driven away from thermodynamic equilibrium often form patterns, if the driving force exceeds a critical value~\cite{cross_hohenberg}. Competing instabilities may lead to interesting pattern dynamics, which helps in understanding the underlying instability mechanism. Several patterns are observed in continuum mechanical systems, such as Rayleigh-B\'enard systems~\cite{RBC_patterns}, B\'enard-Marangoni systems~\cite{benard_marangoni}, magneto-hydrodynamics~\cite{fluid_dynamo}, ferrofluids~\cite{ferro_fluids}, binary fluids~\cite{binary_fluids},  granular materials~\cite{granular_patterns} under shaking, biological systems~\cite{turing_patterns}, etc. Symmetries and dissipation play a very significant role in pattern selection in such systems~\cite{bifurcations}. The selection of a pattern is a consequence of at least one broken symmetry of the system. Unbroken symmetries often introduce multiple patterns, which may lead to a transition from local to global pattern dynamics. The gluing~\cite{gluing} of two limit cycles on two sides of a saddle point in the phase space of a given system is an example of a local to nonlocal bifurcation. It occurs when two limit cycles simultaneously become homoclinic orbits of the same saddle point. This phenomenon has been recently observed in a variety of systems including liquid crystals~\cite{gluing_liquid_crystals}, fluid dynamical systems~\cite{gluing_fluids}, biological systems~\cite{gluing_bioscience}, optical systems~\cite{gluing_optics}, and electrical circuits~\cite{gluing_circuits},  and is a topic of current research. The pattern dynamics in the vicinity of a homoclinic bifurcation has, however, not been investigated in a fluid dynamical system. 

A Rayleigh-B\'enard system~\cite{RBC_books, RBC_review}, where a thin layer of a fluid is heated uniformly from below and cooled uniformly from above, is a classical example of an extended dissipative system which shows a plethora of pattern-forming instabilities~\cite{RBC_patterns}, chaos~\cite{RBC_chaos}, and turbulence~\cite{RBC_turbulence}. Low-Prandtl-number~\cite{RBC_lowP} and very low-Prandtl-number convection~\cite{RBC_zeroP1, RBC_zeroP2} show three-dimensional oscillatory behavior close to the instability onset. In addition, the Rayleigh-B\'enard system  possesses symmetries under translation and rotation in the horizontal plane that can introduce multiple sets of patterns. The possibility of a homoclinic bifurcation and the pattern dynamics in its vicinity are unexplored in three dimensional (3D) Rayleigh-B\'enard convection.

We report, in this article, for the first time the possibility of an inverse homoclinic bifurcation in direct numerical simulations (DNS) of three dimensional (3D) Rayleigh-B\'enard convection (RBC) in low-Prandtl-number fluids, and the results of our investigations of fluid patterns close to the bifurcation. We observe spontaneous breaking of a periodic competition of two mutually perpendicular sets of cross rolls to one set of oscillating cross rolls, as the Rayleigh number $Ra$ is raised above a critical value $Ra_h$. The time period of the oscillating patterns diverges, and shows scaling behavior in the close vicinity of the transition point. The exponents of scaling are asymmetric on the two sides of the transition point. We also present a simple four-mode model, which captures not only the pattern dynamics in the vicinity of the inverse homoclinic bifurcation but also the whole sequence of bifurcations observed in DNS quite well over a wide range of Rayleigh number in low-Prandtl-number fluids. 

The hydrodynamics of RBC in a thin layer of Boussinesq fluid of thickness $d$, kinematic viscosity $\nu$, thermal diffusion coefficient $\kappa$, and thermal expansion coefficient $\alpha$, subjected to an adverse temperature gradient $\beta$,  is governed  by the following dimensionless hydrodynamic equations:
\begin{eqnarray}
\partial_t {\bf v} + ({\bf v \cdot \nabla}){\bf v} &=& -\nabla p + \nabla^2 {\bf v} + Ra\theta{\bf e}_3,\label{eq:velocity}\\
 Pr[\partial_t {\theta} &+&({\bf v}{\cdot}\nabla)\theta]={\nabla}^2 \theta + v_3,\label{eq:theta}\\
\nabla{\cdot}{\bf v} &=& 0, \label{eq:continuity}
\end{eqnarray}
\noindent
where ${\bf v}(x,y,z,t) \equiv (v_1,v_2,v_3)$ is the velocity field, $\theta(x,y,z,t)$ the  convective temperature field, $p$ the pressure due to convection, and ${\bf e}_3 $ a unit vector directed against the direction of the acceleration due to gravity {\bf $g$}. Lengths, time and temperature are measured in units of the fluid thickness $d$, viscous diffusion time $d^2/\nu$, and $\nu \beta d/\kappa$, respectively.  We use thermally conducting and {\it stress-free} boundary conditions which imply that $\theta$ $=$ $ v_3$ $=$ $\partial_zv_1$ $=$ $\partial_zv_2$ $=$ $0$ at $z=0, 1$. All the fields are assumed to be periodic in the horizontal plane. The Rayleigh number $Ra = \alpha \beta g d^4/{\nu\kappa}$ and Prandtl number $Pr = \nu/\kappa$ are two dimensionless numbers that decide the convective flow structures in the fluid. Convection appears when the reduced Rayleigh number $r = Ra/Ra_c$ with $Ra_c = 27\pi^4/4$ is raised above unity. 

We integrate the full hydrodynamic system~(\ref{eq:velocity} - \ref{eq:continuity}) for low-Prandtl-number ($Pr \le 0.025$) fluids using an object oriented code~\cite{verma:Arxiv_2011} based on pseudo-spectral method. The vertical velocity $v_3$ and the temperature field $\theta$ are expanded as: $v_3(x, y, z, t)=\sum_{l, m, n}W_{lmn}(t) e^{ik(l x + m y)} \sin{(n\pi z)}$ and $\theta(x, y, z, t)=\sum_{l, m, n} \Theta_{lmn}(t) e^{ik(l x + m y)} \sin{(n\pi z)}$. The horizontal velocities $v_1$ and $v_2$ have similar expansion in $xy$ plane with co-sinusoidal dependence in the $z$ direction. The integers $l, m, n$ can take values consistent with the equation of continuity (eq.~\ref{eq:continuity}) and $k = k_c = \pi/\sqrt{2}$. The size of the periodic cell for DNS is $2\sqrt{2} \times 2\sqrt{2} \times 1$, and its resolution is $64\times64\times64$. The pattern dynamics is complex at the primary instability in very low-Prandtl-number fluids~\cite{RBC_zeroP2} due to chaotic flows just above the instability onset. We investigate pattern dynamics as soon as we observe the first oscillatory pattern close to the onset of convection. The simulation was started with random initial conditions, and it was continued until a steady state was reached. The steady state  values of fields in a simulation were used as the initial conditions for the next simulation. The value of $r$ was increased in small steps of size $\Delta r$ ($ 0.0001 \le \Delta r \le 0.01$) and the numerical simulations were done for several values of $r$.  We also repeated several runs starting with random initial conditions for different values of $r$ and found no hysteresis in the parameter range considered here. Various observed convective patterns in the DNS for $Pr = 0.01$ and $Pr = 0$ are listed in the first two columns of Table~\ref{table1}. 

\begin{table}[ht]
\caption{Convective patterns computed from DNS in two columns in the middle and from a model in the last column.\\}
\begin{tabular}{|c|c|c|c|}
\hline
 Convective &\multicolumn{2}{|c|}{DNS} & Model \\
\cline {2-4}
patterns & r(Pr = 0.01)& r(Pr = 0) & r(Pr = 0)\\ 
\hline\hline
OCR-I & 1.010 - 1.0835  & 1.0049 - 1.0708 & 1.010 - 1.0953 \\
\hline
OCR-II & 1.0836 - 1.1200 & 1.0709 - 1.1315 & 1.0954 - 1.1584\\
\hline
CR & 1.1210 - 1.1990  & 1.1316 - 1.2005 & 1.1585 - 1.2519\\
\hline
SQ & 1.2000 - 1.4200 & 1.2006 - 1.4297 & 1.2520 - 1.5128\\
\hline
\end{tabular}
\label{table1}
\end{table}

\begin{figure}[h]
\includegraphics[height=!,width=8.6 cm]{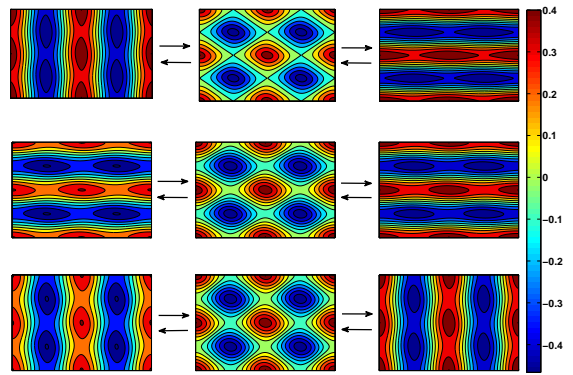}
\caption{Contour plots of the temperature field at the mid plane ($z = 0.5$) near inverse homoclinic bifurcation as observed in DNS ($Pr = 0.01$). The upper row ($r = 1.076$) shows competition between two sets of oscillating cross rolls. The middle and lower rows show  two possibilities of oscillating cross rolls ($r = 1.088$).} \label{fig:DNS_patterns}
\end{figure}

The first ordered state for $Pr =0.01$ appears in the form of a competition of two mutually perpendicular sets of oscillatory cross rolls (OCR-I) which continues to exist until $r = 1.0835$. The first row of Fig.~\ref{fig:DNS_patterns} displays the pattern dynamics for $r = 1.076$. Two sets of cross rolls, one with $|W_{101}| > |W_{011}|$ and another with $|W_{101}| < |W_{011}|$, appear periodically. The patterns appear as squares when the amplitudes of the two sets of cross rolls become equal. The competition represents a global pattern dynamics. As $r$ is raised above a critical $r_h = 1.0835$, the competing cross rolls (OCR-I) spontaneously break into two possible oscillatory cross rolls (OCR-II). The second and third rows of Fig.~\ref{fig:DNS_patterns} show two possibilities of OCR-II for $r = 1.088$. We get oscillating cross-roll patterns with either $|W_{101}| > |W_{011}|$ or $W_{101} < W_{011}$. Two sets of multiple solutions, which are connected by rotation about a vertical axis by $\pi/2$, continue until $r = 1.120$. Further increase in $r$ leads to the appearance of two sets of stationary cross rolls (CR) at $r =1.121$, which is observed till $r = 1.199$.  Raising the value of $r$ even further leads to a transition from stationary cross rolls (CR) to stationary square (SQ) patterns. The similar sequence is observed in the limit of $Pr \rightarrow 0$ (see Table~\ref{table1}). The range of competing cross rolls becomes wider as $Pr$ decreases. We have observed the spontaneous breaking of competing cross rolls to two sets of oscillatory cross rolls in fluids with $0 \le Pr \le 0.025$. 

\begin{figure}[h]
\includegraphics[height=!,width=8.6 cm]{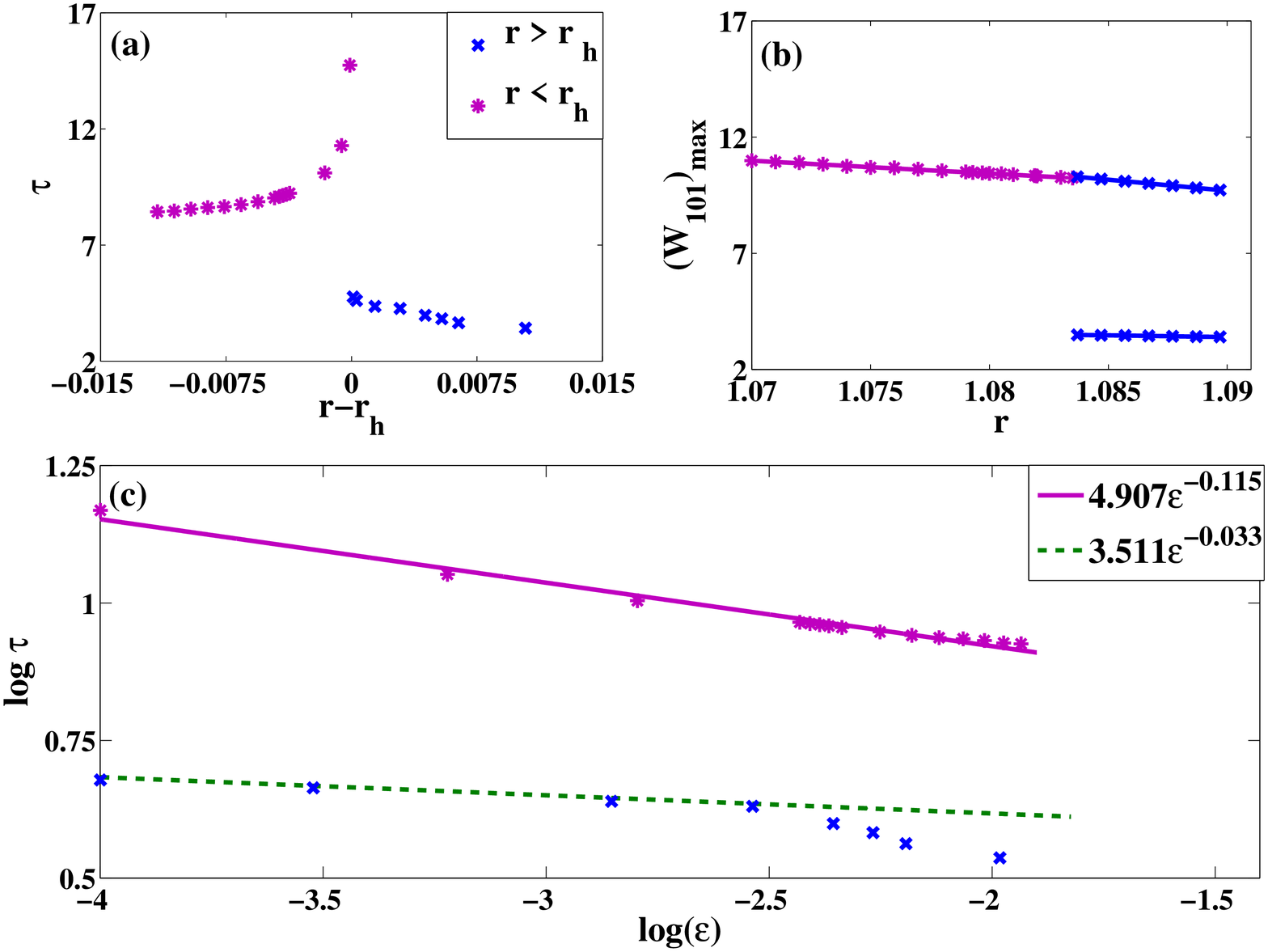}
\caption{Scaling behavior near inverse homoclinic bifurcation ($Pr = 0.01$) as obtained from DNS: (a) The divergence of the dimesionless time period $\tau$ of OCR patterns close to the bifurcation. (b) The variation of maxima of Fourier mode $W_{101}$ with the reduced Rayleigh number $r$ near the bifurcation, showing the spontaneous transition from a global oscillation (star $*$) to two possible local oscillations(cross $\times$) at the homoclinic point. (c) The time period $\tau$ of OCR patterns scales with $\epsilon \equiv |r - r_h|$. The scaling exponents are different before and after the bifurcation.} \label{fig:DNS_gluing}
\end{figure}

Figure~\ref{fig:DNS_gluing} shows the details of a transition from global to local pattern dynamics for $Pr = 0.01$.  The divergence of the time period of oscillatory patterns close to the transition point is displayed in 
Fig.~\ref{fig:DNS_gluing}a. The amplitude of the largest Fourier mode of OCR-I patterns decreases linearly with the increase of $r$ and shows two possible values just above the transition ($r = r_h$) point (Fig.~\ref{fig:DNS_gluing}b). The appearance of two amplitudes signifies a transition from a nonlocal to a local pattern dynamics.  The scaling of time period $\tau$ of oscillating pattern on both sides of the transition point is displayed in Fig.~\ref{fig:DNS_gluing}c, showing asymmetry. The time period $\tau$ of competing patterns scales with $\epsilon\equiv |r - r_h|$ as $\epsilon^{-0.115}$ before transition and as $\epsilon^{-0.033}$ after transition.  The scaling behavior of the time period of OCR suggests the transition to be inverse homoclinic. Simulations~\cite{RBC_zeroP1} of low-$Pr$ RBC with {\it no-slip} boundaries are known to show relaxation oscillation of patterns. The homoclinic instability may be accessible to experiments, if performed in square containers, in this regime.
 
We now construct a simple low dimensional model to analyze pattern dynamics near the inverse homoclinic bifurcation. For this purpose we take the limit of vanishing Prandtl number ($Pr \rightarrow 0$). As the temperature field is slaved to the vertical velocity, the number of modes representing the effective dynamics is expected to be smaller in this limit. We begin with the standard Galerkin technique to derive a low-dimensional-model~\cite{RBC_zeroP2}. We expand the vertical velocity $v_3$ and the vertical vorticity $Z \equiv ({\mathbf \nabla \times v}){\mathbf \cdot e}_3$ such that the essential modes to describe two sets of mutually perpendicular rolls, cross rolls, and the nonlinear interaction between them are retained. We keep five velocity modes $W_{101}$, $W_{011}$, $W_{211}$, $W_{121}$, and $W_{112}$, and two vorticity modes $Z_{110}$ and $Z_{112}$. The hydrodynamic equations are projected on these modes. We then adiabatically  eliminate modes $W_{112}$, $Z_{110}$ and $Z_{112}$. This leads to a simple four-mode model given by,
\begin{eqnarray}
\dot{\bf X} &=& \mu_1{\bf X} +X_1X_2{\Bbb A}(a_{1}{\bf X}+a_{2}{\bf Y})+
               Y_1Y_2{\Bbb A}(a_{3}{\bf X}+a_{4}{\bf Y})               \nonumber\\
	    &+& a_5[X_2^2Y_1, X_1^2Y_2]^{T}+ a_6[X_1Y_2^2, X_2Y_1^2]^{T}, \nonumber\\
\dot{\bf Y} &=& \mu_2{\bf Y} +X_1X_2{\Bbb A}(b_{1}{\bf X}+b_{2}{\bf Y})+
               Y_1Y_2{\Bbb A}(b_{3}{\bf X}+b_{4}{\bf Y})               \nonumber\\
	    &+& b_5[X_2^2Y_1, X_1^2Y_2]^{T}+ b_6[X_1Y_2^2, X_2Y_1^2]^{T},\label{model}
\end{eqnarray}
\noindent 
where ${\bf X}$ $=$ $[X_1, X_2]^T$ $\equiv$ $[W_{101}, W_{011}]^T$, ${\bf Y}$ $=$ $[Y_1, Y_2]^T$ $\equiv$ $[W_{121}, W_{211}]^T$, ${\mathbb A}$ $=$ [$0~~~1;1~~~0$], $\mu_1$ $=$ $3\pi^2(r-1)/2$ and $\mu_2$ $=$  $\pi^2(135r - 343)/98$. The coefficients are: $a_1 = -3/100$, $a_2 = 31/3000$, $a_3 = -209/30000$, $a_4 = 63/60000$, $a_5 = -47/1500$, $a_6 = 1/200$, $b_1 = -93/700$, $b_2 = -67/7000$, $b_3 = -7407/70000$, $b_4 = -3969/700000$, $b_5 = -928/7000$, and $b_6 = 3816/70000$. The superscript $T$ denotes transpose of a matrix. The model is valid for $r < 343/135$ (i.e., $\mu_2 < 0$).

\begin{figure}[h]
\includegraphics[height=8.6 cm,width=8.6 cm]{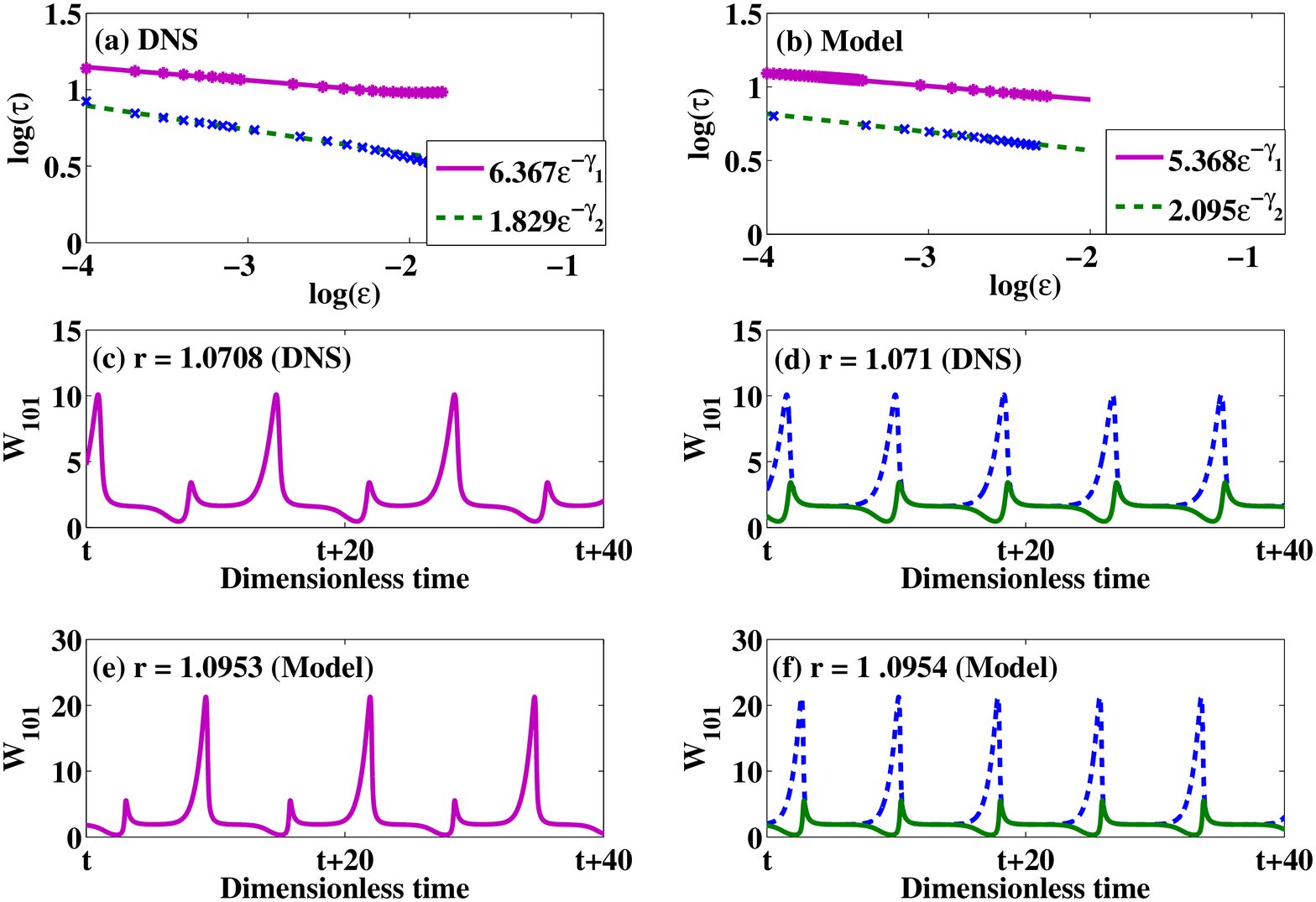} 
\caption{Comparison of results of the model with those from DNS for $Pr=0$: Scaling of the dimensionless time period $\tau$ of the oscillating patterns with $\epsilon \equiv |r-r_h|$ before (solid curve) and after (dashed curve) the bifurcation, as observed in  DNS (a) and the model (b). The exponents from DNS are: $\gamma_1 = 0.086$ and $\gamma_2 = 0.158$, while those from the model are: $\gamma_1 = 0.092$ and $\gamma_2 = 0.124$. Temporal variation of the Fourier mode $W_{101}$ computed from DNS before (c) and after (d) the homoclinic bifurcation. Temporal variation of the mode $W_{101}$ computed from the model before (e) and after (f) the bifurcation.} \label{fig:model_DNS}
\end{figure} 

The model is integrated using standard fourth order Runge Kutta (RK4) method. The second and third columns of Table~\ref{table1} summarize the results obtained from DNS for $Pr =0$ and the model, respectively. The model captures the sequence of bifurcations quite accurately in a wide range of $r$ as observed in DNS. The difference in the lower and the upper bounds for the range of $r$ for any solution computed from the model and DNS is within $6\%$. 

Figure~\ref{fig:model_DNS} gives the comparison of results obtained from the model and those from DNS for $Pr = 0$ near homoclinic bifurcation. The time period $\tau$ of the competing cross rolls scales with $\epsilon \equiv |r-r_h|$ as $\epsilon^{-\gamma_1}$ before the transition (solid curve) and as $\epsilon^{-\gamma_2}$ after the transition (dashed curve). The values of the scaling exponent $\gamma_1$ obtained from the DNS (Fig.~\ref{fig:model_DNS}a) and the model (Fig.~\ref{fig:model_DNS}b) are $0.086$ and $0.092$ respectively. Two exponents are different showing asymmetry in scaling behavior. DNS and the model yield $\gamma_2$ equal to $0.158 $ and $0.124$ respectively.  Figures~\ref{fig:model_DNS}c $\&$ d show the temporal variation of the Fourier mode $W_{101}$ obtained from DNS before and after the bifurcation. The similar behavior is observed in the model (fig.~\ref{fig:model_DNS}e, f). The model reveals that the unstable $SQ$ patterns exist as saddle fixed points for $1 \le r < 1.252$ but become stable at $r = 1.252$. The competing cross rolls (OCR-I) break into two possible sets of OCR-II when the amplitude of OCR-I oscillation touches a saddle square. This confirms the transition from OCR-I to OCR-II as an inverse homoclinic bifurcation. The model includes very few modes, and therefore shows higher values of the Fourier mode $W_{101}$. The time periods of oscillating patterns obtained from the model and DNS are in good agreement both before and after the transition. 

We have investigated the possibility of an inverse homoclinic bifurcation and associated pattern dynamics in RBC.  The spontaneous breaking of a competition between two mutually perpendicular sets of oscillatory cross rolls into one set of oscillatory cross rolls occurs close to the bifurcation point. The time period of oscillation shows asymmetric scaling on the two sides of the bifurcation point. We have also constructed a simple four-mode model which captures accurately the sequence of bifurcations including the pattern dynamics near the homoclinic bifurcation. The model with different values of the coefficients $a_i$ and $b_i$ may be useful to study pattern dynamics on square lattices in other extended systems with similar symmetries. 

We have benefitted from fruitful discussions with J.K. Bhattacharjee, H. Pharasi, L.K. Dey, and D. Kumar.

\end{document}